\documentclass[prd,
eqsecnum,floatfix,letterpaper,superscriptaddress,groupedaddress,nofootinbib]{revtex4}

\usepackage{amsmath}
\usepackage{latexsym}
\usepackage{amssymb}
\usepackage{amsfonts}
\usepackage{mathtools}
\usepackage{bm}
\usepackage{amsthm}
\usepackage{graphicx}
\usepackage{color}
\usepackage{changes}
\usepackage[normalem]{ulem}
\usepackage{natbib}
\usepackage{appendix}
\def \R {{\mathcal R}}
\def\no{\nonumber}
\def \D{{\mathcal D}}
\def \G{{\cal G}}
\def \a {\alpha}
\def \b {\beta}
\def \tpp {t^{\prime \prime}}

\def \be{\begin{equation}}
\def \bea{\begin{eqnarray}}
\def \eea{\end{eqnarray}}
\def \ee{\end{equation}}

\newtheorem{mydef}{Definition}

\begin{document}
\title{Matrix Representation of Time-Delay Interferometry}
\author{Massimo Tinto}
\email{mtinto@ucsd.edu}
\affiliation{University of California San Diego,
  Center for Astrophysics and Space Sciences,
  9500 Gilman Dr, La Jolla, CA 92093,
  U.S.A.}
\affiliation{Divis\~{a}o de Astrof\'{i}sica, Instituto
  Nacional de Pesquisas Espaciais, S. J. Campos, SP 12227-010, Brazil}
\author{Sanjeev Dhurandhar}
\email{sanjeev@iucaa.in}
\affiliation{Inter University Centre for Astronomy and Astrophysics,
  Ganeshkhind, Pune, 411 007, India}
\author{Prasanna Joshi}
\email{prasanna.mohan.joshi@aei.mpg.de}
\affiliation{Max Planck Institute for Gravitational Physics (Albert-Einstein-Institute), D-30167 Hannover, Germany}

\date{\today}

\begin{abstract}
  Time-Delay Interferometry (TDI) is the data processing technique
  that cancels the large laser phase fluctuations affecting the
  one-way Doppler measurements made by unequal-arm space-based
  gravitational wave interferometers. By taking finite linear
  combinations of properly time-shifted Doppler measurements, laser
  phase fluctuations are removed at any time $t$ and gravitational
  wave signals can be studied at a requisite level of precision.

  In this article we show the delay operators used in TDI can be
  represented as matrices acting on arrays associated with the laser
  noises and Doppler measurements. The matrix formulation is nothing
  but the group theoretic representation (ring homomorphism) of the
  earlier approach involving time-delay operators and so in principle
  is the same. It is shown that the homomorphism is valid generally and  we cover all situations of interest. To understand the potential
  advantages the matrix representation brings, care must be taken by
  the data analyst to account for the light travel times when linearly
  relating the one-way Doppler measurements to the laser noises. This
  is especially important in view of the future gravitational wave
  projects envisaged. We show that the matrix formulation of TDI
  results in the cancellation of the laser noises at an arbitrary time
  $t$ by only linearly combining a finite number of samples of the
  one-way Doppler data measured at and around time $t$.
\end{abstract}

\maketitle

\section{Introduction}
\label{SecI}

Interferometric detectors of gravitational waves with frequency
content $0 < f < f_0$ may be thought of as optical configurations with
one or more arms folding coherent trains of electromagnetic waves (or
beams) of nominal frequency $\nu_0 \gg f_0$. At points where these
intersect, relative fluctuations of frequency or phase are monitored
(homodyne detection).  Frequency fluctuations in a narrow Fourier band
can alternatively be described as fluctuating side-band
amplitudes. Interference of two or more beams, produced and monitored
by a (nonlinear) device such as a photo detector, exhibits these
side-bands as a low frequency signal again with frequency content
$0 < f < f_0$.  The observed low frequency signal is due to frequency
variations of the sources of the beams about $\nu_0$, to relative
motions of the sources and any mirrors (or amplifying microwave or
optical transponders) that do any beam folding, to temporal variations
of the index of refraction along the beams, and, according to general
relativity, to any time-variable gravitational fields present, such as
the transverse trace-less metric curvature of a passing plane
gravitational wave train.  To observe these gravitational fields in
this way, it is thus necessary to control, or monitor, the other
sources of relative frequency fluctuations, and, in the data analysis,
to optimally use algorithms based on the different characteristic
interferometer responses to gravitational waves (the signal) and to
the other sources (the noise).

By comparing phases of split beams propagated along equal but
non-parallel arms, frequency fluctuations from the source of the beams
are removed directly at the photo detector and gravitational wave
signals at levels many orders of magnitude lower can be detected.
Especially for interferometers that use light generated by presently
available lasers, which display frequency stability roughly a few
parts in $10^{-13}$ in the millihertz band, it is essential to remove
these fluctuations when searching for gravitational waves of
dimensionless amplitude smaller than $10^{-21}$.

Space-based, two-arm interferometers
\cite{LISA2017,Taiji,TianQin,gLISA2015,Astrod} are prevented from
canceling the laser noise by directly interfering the beams from the
two unequal arms at a single photo detector because laser phase
fluctuations experience different delays. As a result, two Doppler
data from the two arms are measured at two different photo detectors
and are then digitally processed to compensate for the inequality of
the arms. This data processing technique, called Time-Delay
Interferometry (TDI) \cite{TD2020}, entails time-shifting and linearly
combining the two Doppler measurements so as to achieve the required
sensitivity to gravitational radiation.

In a recent article \cite{Vallisneri2020}, a data processing
alternative to TDI has been proposed for the two-arm
configuration. This technique, which has been named TDI-$\infty$ (as
it cancels the laser noise at an arbitrary time $t$ by linearly
combining all the Doppler measurements made up to time $t$), relies on
an identified linear relationship between the two Doppler measurements
made by an unequal-arm Michelson interferometer and the laser
noise. Based on this formulation, TDI-$\infty$ cancels laser phase
fluctuations by applying linear algebra manipulations to the Doppler
data. Through its implementation, TDI-$\infty$ is claimed to (i)
simplify the data processing for gravitational wave signal searches in
the laser-noise-free data over that of TDI, (ii) work for any
time-dependent light-time delays, and (iii) automatically handle data
gaps.

After briefly reviewing the TDI-$\infty$ technique for the two
unequal-arm configuration, we show care must be taken to account for
the light-travel-times when linearly relating the two-way Doppler
measurements to the laser noise \cite{Vallisneri2020}. The two-way
Doppler data at a time $t$ is the result of the interference between
the returning beam and the outgoing beam. As such it contains the
difference between the value of the laser noise at time $t - l_i(t)$
affecting the returning beam (with $l_i(t)$ being the
round-trip-light-time (RTLT)) and the laser noise of the outgoing beam
at time $t$ when the measurement is recorded. From the instant the
laser is switched on (let us say $t=0$) each two-way Doppler
measurement becomes different from zero only for $t \ge l_i (t)$,
i.e. when the returning beam and the outgoing beam start to
interfere. By accounting for this observation in the ``boundary
conditions'' of the Doppler data,  we show that it is
  possible to introduce a matrix representation of TDI.

We would like to briefly mention here another matrix based
approach. Romano and Woan \cite{PCA2006} have used Bayesian inference
to set up a noise covariance matrix of the data streams. Then by
performing a principal component analysis of the covariance matrix,
they identify the principal components with large eigenvalues with the
laser noise and so distinguish it from other ambient noises and signal
which correspond to small eigenvalues. We argue that this approach is
also a matrix representation of the original TDI.

Here we provide a summary of this article. In section \ref{SecII} we
present the key-points of TDI-$\infty$ and correct the expression of
the matrix introduced in \cite{Vallisneri2020} relating the two arrays
associated with the two-way Doppler measurements to the array of the
laser noise. We then recast this linear relationship in terms of two
square-matrices, each relating the array associated with one of the
two-way Doppler measurement to the array of the laser noise. As
expected these matrices are singular, reflecting the physical
impossibility of reconstructing the laser noise array from the arrays
associated with the two-way Doppler data. In the simple configuration
of a stationary interferometer whose RTLTs are integer-multiples of
the sampling time, we show that the linear combination of the two-way
Doppler arrays canceling the laser noise is equal to the sampled
unequal-arm Michelson TDI combination $X$. In section \ref{SecIII} we
then turn to the problem of a stationary three-arm array with three
laser noises and six one-way Doppler measurements. After deriving the
expressions of the matrices relating the laser noises to the one-way
Doppler measurements, we show that the generators of the space of the
combinations canceling the laser noises are equal to the sampled
TDI-combinations ($\alpha, \beta, \gamma, \zeta$) \cite{TD2020} in
which the delay operators have been replaced by our derived
matrices. This is rigorously established in section \ref{represent} by
showing that the matrix formulation is just a ring representation of
the first module of syzygies - a ring homomorphism. We cover all cases of interest. We first start with delays that are integer multiples of the
sampling interval, then the continuum case when the sampling is continuous
and the sampling interval tends to zero and finally when fractional-delay filtering based on Lagrange polynomials is used for reconstructing the samples at any required time. 
For fractional delays we show that homomorphism is valid (i) when all 
delays lie in the same interpolation interval, (ii) for each delay lying in different interpolation intervals and
also (iii) for time-dependent arm-lengths. In all these cases we show that there
is a ring homomorphism. Thus the matrix formulation is in principle
the same as the original formulation of TDI, although it might offer
some advantages when implemented numerically. Finally, in section
\ref{SecIV} we present our concluding remarks and summarize our
thoughts about potential advantages in processing the TDI measurements
cast in matrix form when searching for gravitational wave signals.

\section{Matrix representation of the two-way Doppler measurements}
\label{SecII}

Equal-arm interferometer detectors of gravitational waves can observe
gravitational radiation by canceling the laser frequency fluctuations
affecting the light injected into their arms. This is done by
comparing phases of split beams propagated along the equal (but
non-parallel) arms of the detector. The laser frequency fluctuations
affecting the two beams experience the same delay within the two
equal-length arms and cancel out at the photo-detector where relative
phases are measured. This way gravitational-wave signals of
dimensionless amplitude less than $10^{-22}$ can be observed when
using lasers whose frequency stability can be as large as roughly a
few parts in $10^{-18}$ in the kilohertz band.

If the arms of the interferometer have different lengths, however, the
exact cancellation of the laser frequency fluctuations, say $C (t)$,
will no longer take place at the photo-detector. In fact, the larger
the difference between the two arms, the larger will be the magnitude
of the laser frequency fluctuations affecting the detector response.
If $l_1$ and $l_2$ are the RTLTs of the laser beams
within the two arms, it is easy to see that the amount of laser
relative frequency fluctuations remaining in the response are equal
to:
\begin{equation}
  \Delta C (t) = C(t - l_1) - C(t - l_2) \, .
  \label{DC}
\end{equation}
In the case of a space-based interferometer such as LISA for instance,
whose lasers are expected to display relative frequency fluctuations
equal to about $10^{-13}/\sqrt{\mathrm{Hz}}$ in the mHz band and
RTLTs will differ by a few percent \cite{LISA2017},
Eq.~(\ref{DC}) implies uncanceled fluctuations from the laser as large
as $\approx 10^{-16}/\sqrt{\mathrm{Hz}}$ at a millihertz frequency
\cite{TD2020}. Since the LISA sensitivity goal is about
$10^{-20}/\sqrt{\mathrm{Hz}}$ in this part of the frequency band, it
is clear that an alternative experimental approach for canceling the
laser frequency fluctuations is needed.

\begin{figure}[!ht]
\begin{center}
\includegraphics[width = \linewidth, clip]{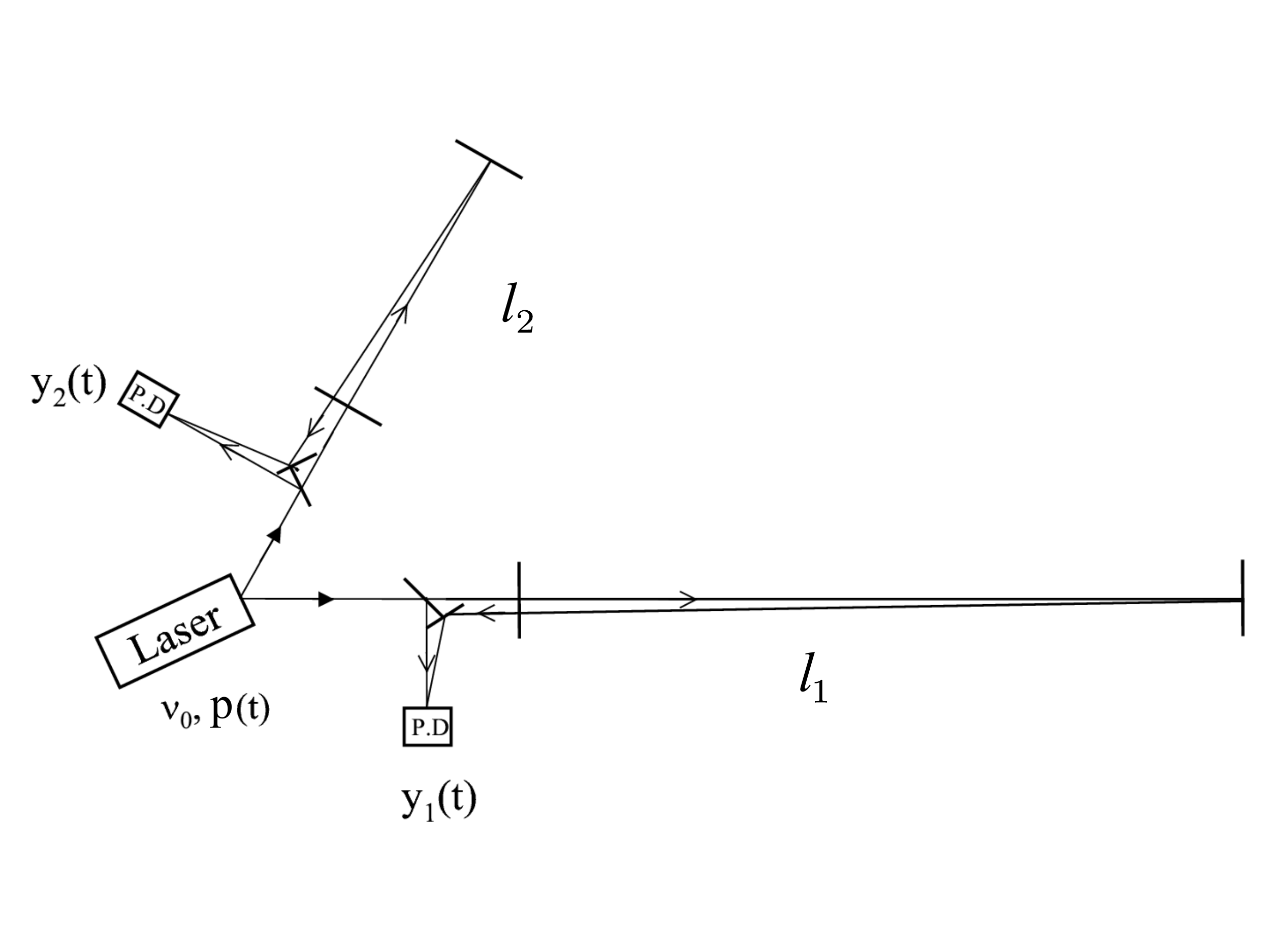}
\end{center}
\caption{Light from a laser is split into two beams, each
      injected into an arm formed by pairs of free-falling
      mirrors. Since the RTLTs, $l_1$ and $l_2$, are
      different, now the light beams from the two arms are not
      recombined at one photo detector. Instead each is separately made
      to interfere with the light that is injected into the arms. Two
      distinct photo detectors are now used, and phase (or frequency)
      fluctuations are then monitored and recorded there.}
\label{Fig0}
\end{figure}

An elegant method entirely implemented in the time domain was first
suggested in \cite{TA99} and then generalized in a series of related
publications (see \cite{TD2020} and references therein). Such a
method, named Time-Delay Interferometry (or TDI) as it requires
time-shifting and linearly combining the recorded data, carefully
accounts for the time-signature of the laser noise in the two-way
Doppler data. TDI relies on the optical configuration exemplified by
Fig.~\ref{Fig0} \cite{FB84,FBHHV85, FBHHSV89,GHTF96}. In this
idealized model the two beams exiting the two arms are
{\underbar{not}} made to interfere at a common photo-detector. Rather,
each is made to interfere with the incoming light from the laser at a
photo-detector, decoupling in this way the laser phase fluctuations
experienced by the two beams in the two arms. In the case of a
stationary array, cancellation of the laser noise at an arbitrary time
$t$ requires only four samples of the measurements made at and around
$t$. Contrary to a previously proposed technique \cite{FB84,FBHHV85,
  FBHHSV89,GHTF96}, which required processing in the Fourier domain a
large ($\sim$ six months) amount of data to sufficiently suppress the
laser noise at a time $t$ \cite{TA99,TD2020}, TDI can be regarded as a
``local'' method.

In a recent publication \cite{Vallisneri2020}, a new ``global''
technique for canceling the laser noise has been proposed. This
technique, which has been named TDI-$\infty$, establishes a linear
relationship between the sampled Doppler measurements and the laser
noise arrays. It is claimed to work for any time-dependent delays and
to cancel the laser noise at an arbitrary time $t$ by taking linear
combinations of the two-way Doppler measurements sampled at all times
before $t$.

To understand the formulation of TDI-$\infty$, let us consider again
the simplified (and stationary) two-arm optical configuration shown in
Figure \ref{Fig0}.  In it the laser noise, $C(t)$, folds into the two
two-way Doppler data, $y_1(t)$, $y_2(t)$, in the following way (where
we disregard the contributions from all other physical effects
affecting the two-way Doppler data):
\begin{eqnarray}
  y_1(t) =  C(t - l_1(t)) - C(t)  \ ,
\nonumber
  \\
  y_2(t) = C(t - l_2(t)) - C(t)  \ ,
\label{eq1}
\end{eqnarray}
where $l_1$, $l_2$ are the two RTLTs, in general also
functions of time $t$.

\begin{figure}
\centering
\includegraphics[width = 7in]{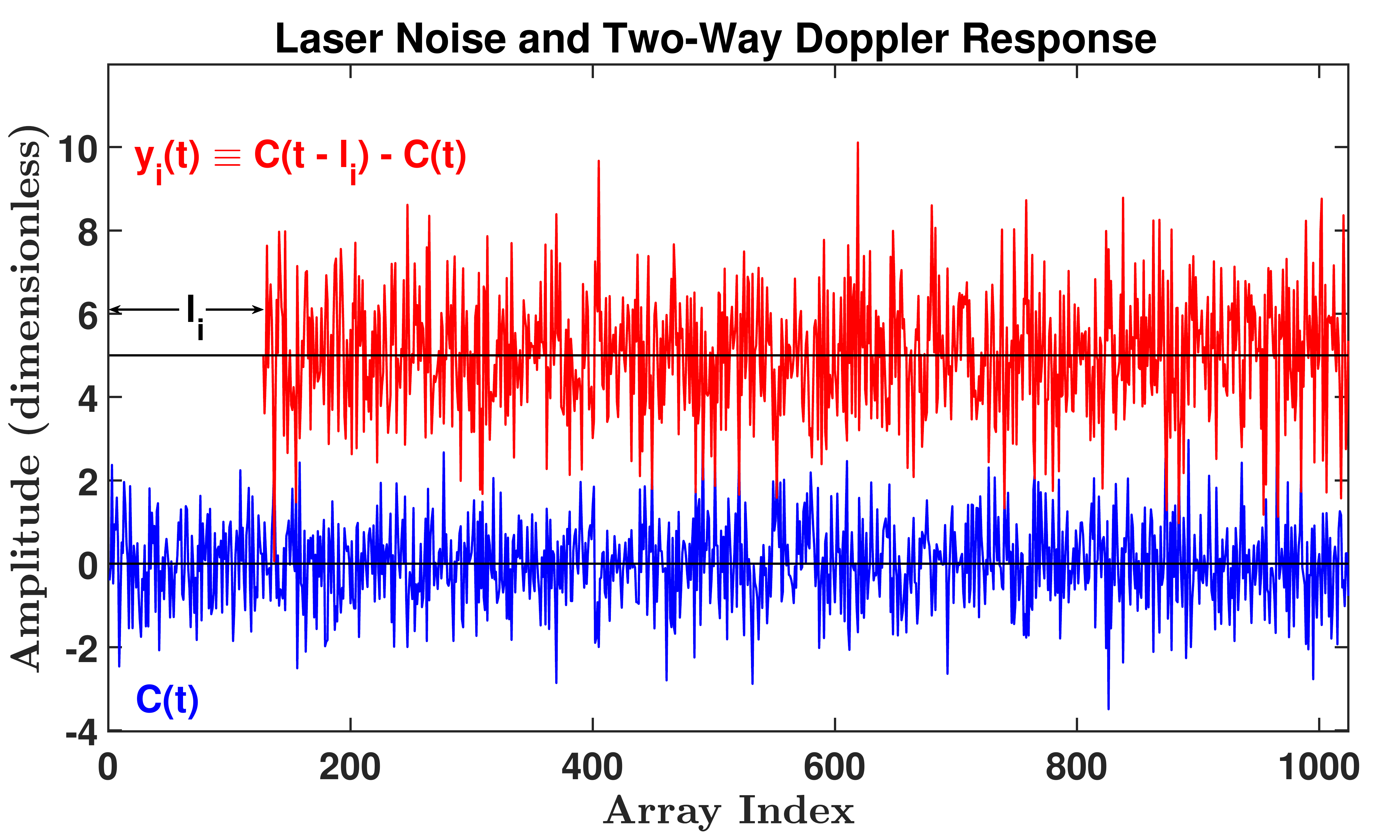}
\caption{The laser noise random process (blue color), $C(t)$, together
  with the corresponding two-way Doppler measurement (red color),
  $y_i$. The laser is switched on at time $t=0$. Since it takes $l_i$
  seconds for the beam to return to the transmitting spacecraft, 
  $y_i$ is identically equal to zero since no Doppler measurements can
  be performed during this time interval.}
\label{fig1}
\end{figure}

Operationally, Eq. (\ref{eq1}) says that each sample of the two-way
Doppler data at time $t$ contains the difference between the laser
noise $C$ generated at a RTLT earlier, $t - l_i(t) \ , \ i = 1, 2$ and
that generated at time $t$. Figure \ref{fig1} displays graphically
what we have just described. The important point to note here is what
happens during the first $l_i$ seconds from the instant $t = 0$ when
the laser is switched on. Since the $y_i$ measurements are the result
of interfering the returned beam with the outgoing one, during the
first $l_i$ seconds (i.e. from the moment the laser has been turned
on) the $y_i$ measurements are identically equal to zero because no
interference measurements can be performed during this time. In other
words, during the first $l_i$ seconds there is not yet a returning
beam with which the local light is made to interfere with. In
\cite{Vallisneri2020}, however, only the first terms on the
right-hand-sides of Eq. (\ref{eq1}) were disregarded during these time
intervals. Although the TDI-$\infty$ technique is mathematically
correct, by using these nonphysical "boundary conditions" results in
solutions that do not cancel the laser noise when applied to the
Doppler data measured by future space-based interferometers. We
verified this analytically (by implementing the TDI-$\infty$ algorithm
with the help of the program {\it Mathematica} \cite{Wolf02}) when the
two light-times are constant and equal to integer multiples of the
sampling time. We found the resulting solutions to be linear
combinations of the TDI unequal-arm Michelson combination $X$ defined
at each of the sampled times, plus \textit{an additional term} that
would not cancel the laser noise in the measured data. This additional
term is a function of $y_1$ and $y_2$ defined at times $t < l_1, l_2$
and thus, is a manifestation of the non-physical boundary conditions.
In the attempt of avoiding this problem one might consider start
processing the Doppler data at any time $t$ after the first RTLT has
past. However, one would still be confronted by the fact that the
Doppler measurement $y_i$ at time $t$ contains laser noise generated
at time $t$ and at time $t - l_i$. In other words, there exists a
time-mismatch between the array of the Doppler measurement and that of
the laser noise and physical boundary conditions have to be accounted
for in a realistic simulation. 

TDI-$\infty$ is a ``global'' data processing algorithm, i.e. its
solutions at time $t$ require use of all samples of the Doppler
measurements recorded up to time $t$. Our computations for assessing
the effects of the nonphysical boundary conditions were carried out
only for time intervals relatively short, namely, for stretches of
data containing about 200 samples. Although they indicate the
dependence of the solutions on the boundary conditions, it is possible
that for year-long stretches of data the effects of the selected
boundary conditions might not be significant. This, however, needs to
be mathematically proved. A detailed mathematical investigation of
this point should be carried out in the future and may require
extensive work.

In TDI-$\infty$ the sampled two two-way Doppler data are
packaged in a single array in an alternating fashion starting from
time $t=t_0$ when the laser is switched on. Assuming a stationary
array configuration in which the RTLTs $l_1$, $l_2$ are equal to twice
and three times the sampling time $\Delta t$, (as exemplified in
\cite{Vallisneri2020}), the measurements array is linearly related to
the array associated with the samples of the laser noise $C$ through a
rectangular $2N \times N$ matrix $M$ ($N$ being the number of
considered samples) in the following way:
\begin{equation}
\left(
    \begin{array}{c}
        y_1(t_0) \\
        y_2(t_0) \\
        y_1(t_1) \\
        y_2(t_1) \\
        y_1(t_2) \\
        y_2(t_2) \\
        y_1(t_3) \\
        y_2(t_3) \\
        y_1(t_4) \\
        y_2(t_4) \\
        \vdots
    \end{array}
\right)
=
\left(
    \begin{array}{rrrrrr}
         -1 &  0 &  0 & 0  & 0  & \cdots \\
         -1 &  0 &  0 & 0  & 0  & \cdots \\
         0 &  -1 &  0 & 0  & 0  & \cdots \\
         0 &  -1 &  0 & 0  & 0  & \cdots \\
         1 &  0 &  -1 & 0  & 0  & \cdots \\
         0 &  0 &  -1 & 0  & 0  & \cdots \\
         0 &  1 &  0 & -1 & 0  & \cdots \\
         1 &  0 &  0 & -1 & 0  & \cdots \\
         0 &  0 &  1 &  0 & -1 & \cdots \\
         0 &  1 &  0 &  0 & -1 & \cdots \\
         \vdots & \vdots & \vdots & \vdots & \vdots & \ddots
    \end{array}
\right)
\cdot
\left(
    \begin{array}{c}
        c(t_0) \\
        c(t_1) \\
        c(t_2) \\
        c(t_3) \\
        c(t_4) \\
        \vdots
    \end{array}
\right) \ .
\label{design}
\end{equation}
As shown by Eq. (\ref{design}), rows $1$ through $4$ and row $6$ of
matrix $M$ reflect the assumption made in \cite{Vallisneri2020} of the
two Doppler measurements to contain the laser noise $C$ only at time
$t$ during the time intervals $t_0 \le t < t_0 + l_i$. If, on the
other hand, we correctly assume rows $1$ through $4$ and row $6$ to be
identically equal to zero, the null-space associated to the matrix $M$
will clearly be different.

To better understand and quantify this difference, we split the above
measurement's array in two arrays, ($Y_1, Y_2$), (one per measurement)
and introduce two corresponding ($N \times N$) square-matrices
relating the measurement arrays to the array of the laser noise.  We
assume again a stationary configuration with RTLTs ($l_1, l_2$) equal
to twice and three-times the sampling time respectively. The two
vectors, $Y_1$, $Y_2$, are related to the laser noise vector $C$
through the following expressions:
\begin{equation}
  Y_1 = A_1 . C \ \ \ ; \ \ \ Y_2 = A_2 . C \ ,
  \label{eq5}
\end{equation}
where the symbol $.$ denotes matrix multiplication, and $A_1$, $A_2$
are equal to the following square-matrices:
\begin{equation}
A_1 = \left(
    \begin{array}{rrrrrrr}
         0 &  0 &  0 &  0 & 0 & 0 & \cdots \\
         0 &  0 &  0 &  0 & 0  & 0 & \cdots \\
         1 &  0 &  -1 &  0 & 0 & 0  & \cdots \\
         0 &  1 &  0 &  -1 & 0 & 0 & \cdots \\
         0 &  0 &  1 &  0 & -1 & 0 & \cdots \\
         0 &  0 &  0 &  1 & 0 &  & -1 \cdots \\
         \vdots & \vdots & \vdots & \vdots & \vdots & \ddots & \ddots
    \end{array}
  \right) \ \ \ \ ,
A_2 = \left(
    \begin{array}{rrrrrrr}
         0 &  0 &  0 &  0 & 0 & 0  & \cdots \\
         0 &  0 &  0 &  0 & 0 & 0  & \cdots \\
         0 &  0 &  0 &  0 & 0 & 0  & \cdots \\
         1 &  0 &  0 &  -1 & 0 & 0 & \cdots \\
         0 &  1 &  0 &  0 & -1 & 0 & \cdots \\
         0 &  0 &  1 &  0 & 0 & -1 & \cdots \\
         \vdots & \vdots & \vdots & \vdots & \vdots & \ddots & \ddots
    \end{array}
  \right) \ .
\end{equation}
Note the above matrices incorporate the correct ``boundary
conditions'' as their first few rows are null (the number of null rows
depends on the magnitude of the RTLT).  It is evident that the rank of
the matrices $A_1$, $A_2$ is less than the number of samples $N$ and
therefore they cannot be inverted. Physically this means that,
although the laser noise cannot be known/measured at any time $t$, one
can still cancel it by taking suitable linear combinations of the
two-way Doppler data. Let us consider the following linear combination
of the two-way Doppler measurements:
\begin{equation}
  X \equiv A_2 . Y_1 - A_1 . Y_2 = (A_2 . A_1 - A_1 . A_2) . C \ .
\end{equation}
We have verified that the commutator $[A_1, A_2]$ starts to become
zero from row 6 onward.  If we write the vectors $Y_1$, $Y_2$ in terms
of their components, the linear combination $X$ becomes equal to:
\begin{equation}
  X = 
\left(
    \begin{array}{c}
        0 \\
        0 \\
        0 \\
        y_1(t_3) - y_2(t_3) \\
        y_1(t_4) - y_2(t_4) \\
        -y_1(t_2) + y_1(t_5) + y_2(t_3) - y_2(t_5) \\
        -y_1(t_3) + y_1(t_6) + y_2(t_4) - y_2(t_6) \\
        \vdots
    \end{array}
\right)
\label{X}
\end{equation}
The above vector $X$ is no other than the unequal-arm Michelson TDI
combination sampled at successive sampling times. Note that $X$ starts
to cancel the laser noise after $l_1 + l_2 = 5 \Delta t$ time-samples
have past.

If we would incorporate in the matrices $A_1$, $A_2$ nonphysical ``boundary conditions'', they now would be of rank $N$
and therefore invertible. As each Doppler data could be used to
reconstruct the laser noise, one could then derive a laser-noise-free
combination sensitive to gravitational radiation by taking the
difference of the two reconstructions. Since each reconstruction of
the laser noise at time $t$ would be a linear combination of samples
taken at times determined only by the RTLT of the time-series used,
any time-dependence of the RTLT could be accommodated. In other words,
since each time-series would not be delayed by the RTLT associated
with the other time-series, issues related to the non-commutativity of
the delay operators would not be present.

\section{Matrix Formulation of TDI}
\label{SecIII}

In the general case of three arms, we have six one-way Doppler
measurements and three independent laser noises. The analysis below
will also assume a stationary array and the one-way-light-times to be
equal to $L_1 = \Delta t$, $L_2 = 2 \ \Delta t$, $L_3 = 3 \ \Delta t$
respectively. Although these RTLTs do not reflect the array's
triangular shape, we adopt them so that we can minimize the size of
the matrices introduced for explaining our method without loss of
generality.

By generalizing what was described in the previous section for the $X$
combination, we may write the one-way Doppler data in terms of the
laser noises in the following form (using the notation introduced in
\cite{TD2020}):
\begin{eqnarray}
y_1 & = & D_3 . C_2 - I_3 . C_1 \ \ , \ \   y_{1'} = D_2 .  C_3 - I_2 . C_1
\nonumber
\\
y_2 & = & D_1 . C_3 - I_1 . C_2 \ \ , \ \   y_{2'} = D_3 .  C_1 - I_3 . C_2
\nonumber
\\
y_3 & = & D_2 . C_1 - I_2 . C_3 \ \ , \ \   y_{3'} = D_1 .  C_2 - I_1 . C_3
\label{oneways}
\end{eqnarray}
In Eq.(\ref{oneways}) we index the one-way Doppler data as
  follows: the beam arriving at spacecraft $i$ has subscript $i$ and
  is primed or unprimed depending on whether the beam is traveling
  clockwise or counter-clockwise around the interferometer array, with
  the sense defined by a chosen orientation of the array; the
matrices $D_{i}$ correspond to the delay operators of TDI and there
are only three of them because the array is stationary
\cite{TD2020}. The expressions for $D_{i}$ and $I_{i}$ are equal to:
\begin{equation}
D_3 = \left(
    \begin{array}{rrrrrrrr}
         0 &  0 &  0 &  0 & 0  & 0 & 0 & \cdots \\
         0 &  0 &  0 &  0 & 0  & 0 &  0 & \cdots \\
         0 &  0 &  0 &  0 & 0  & 0 &  0 & \cdots \\
         1 &  0 &  0 &  0 & 0 & 0 &  0 & \cdots \\
         0 &  1 &  0 &  0 & 0 & 0 &  0 & \cdots \\
         0 &  0 &  1 &  0 & 0 & 0 &  0 & \cdots \\
         0 &  0 &  0 &  1 & 0 & 0 &  0 & \cdots \\
\vdots & \vdots & \vdots & \vdots & \vdots & \vdots  & \vdots & \ddots
    \end{array}
  \right) \ \ \ \ ,
I_3 = \left(
     \begin{array}{rrrrrrrr}
         0 &  0 &  0 &  0 & 0  & 0 & 0 & \cdots \\
         0 &  0 &  0 &  0 & 0  & 0 &  0 & \cdots \\
         0 &  0 &  0 &  0 & 0  & 0 &  0 & \cdots \\
         0 &  0 &  0 &  1 & 0 & 0 &  0 & \cdots \\
         0 &  0 &  0 &  0 & 1 & 0 &  0 & \cdots \\
         0 &  0 &  0 &  0 & 0 & 1 &  0 & \cdots \\
         0 &  0 &  0 &  0 & 0 & 0 &  1 & \cdots \\
\vdots & \vdots & \vdots & \vdots & \vdots & \vdots  & \vdots & \ddots
    \end{array}
  \right) \ ;
  \label{op1}
\end{equation}

\begin{equation}
D_2 = \left(
    \begin{array}{rrrrrrrr}
         0 &  0 &  0 &  0 & 0  & 0 & 0 & \cdots \\
         0 &  0 &  0 &  0 & 0  & 0 &  0 & \cdots \\
         1 &  0 &  0 &  0 & 0  & 0 &  0 & \cdots \\
         0 &  1 &  0 &  0 & 0 & 0 &  0 & \cdots \\
         0 &  0 &  1 &  0 & 0 & 0 &  0 & \cdots \\
         0 &  0 &  0 &  1 & 0 & 0 &  0 & \cdots \\
         0 &  0 &  0 &  0 & 1 & 0 &  0 & \cdots \\
\vdots & \vdots & \vdots & \vdots & \vdots & \vdots  & \vdots & \ddots
    \end{array}
  \right) \ \ \ \ ,
I_2 = \left(
     \begin{array}{rrrrrrrr}
         0 &  0 &  0 &  0 & 0  & 0 & 0 & \cdots \\
         0 &  0 &  0 &  0 & 0  & 0 &  0 & \cdots \\
         0 &  0 &  1 &  0 & 0  & 0 &  0 & \cdots \\
         0 &  0 &  0 &  1 & 0 & 0 &  0 & \cdots \\
         0 &  0 &  0 &  0 & 1 & 0 &  0 & \cdots \\
         0 &  0 &  0 &  0 & 0 & 1 &  0 & \cdots \\
         0 &  0 &  0 &  0 & 0 & 0 &  1 & \cdots \\
\vdots & \vdots & \vdots & \vdots & \vdots & \vdots  & \vdots & \ddots
    \end{array}
  \right) \ ;
  \label{op2}
\end{equation}

\begin{equation}
D_1 = \left(
    \begin{array}{rrrrrrrr}
         0 &  0 &  0 &  0 & 0  & 0 & 0 & \cdots \\
         1 &  0 &  0 &  0 & 0  & 0 &  0 & \cdots \\
         0 &  1 &  0 &  0 & 0  & 0 &  0 & \cdots \\
         0 &  0 &  1 &  0 & 0 & 0 &  0 & \cdots \\
         0 &  0 &  0 &  1 & 0 & 0 &  0 & \cdots \\
         0 &  0 &  0 &  0 & 1 & 0 &  0 & \cdots \\
         0 &  0 &  0 &  0 & 0 & 1 &  0 & \cdots \\
         \vdots & \vdots & \vdots & \vdots & \vdots & \vdots  & \vdots & \ddots
    \end{array}
  \right) \ \ \ \ ,
I_1 = \left(
     \begin{array}{rrrrrrrr}
         0 &  0 &  0 &  0 & 0  & 0 & 0 & \cdots \\
         0 &  1 &  0 &  0 & 0  & 0 &  0 & \cdots \\
         0 &  0 &  1 &  0 & 0  & 0 &  0 & \cdots \\
         0 &  0 &  0 &  1 & 0 & 0 &  0 & \cdots \\
         0 &  0 &  0 &  0 & 1 & 0 &  0 & \cdots \\
         0 &  0 &  0 &  0 & 0 & 1 &  0 & \cdots \\
         0 &  0 &  0 &  0 & 0 & 0 &  1 & \cdots \\
\vdots & \vdots & \vdots & \vdots & \vdots & \vdots  & \vdots & \ddots
    \end{array}
  \right) \ .
  \label{op3}
\end{equation}

The problem of identifying all possible TDI combinations associated with the
six one-way Doppler measurements becomes one of determining six
matrices, $q_{i} , q_{i'}$ such that the following equation holds:
\begin{equation}
\sum_{i=1}^{3} q_i . y_i + \sum_{i'=1}^{3} q_{i'} . y_{i'} = 0 \ ,
  \label{TDI}
\end{equation}
where the above equality means ``zero laser noises''. Before
proceeding, note that the matrices $D_i$ and $I_i$ satisfy the
following identities which may be
useful later on:
\bea
I_i . D_k &=& D_k \,, ~~~~~ i \leq k
\no \\
&\neq& D_k \,, ~~~~~~ i > k
\no \\
I_i . I_k &=& I_k . I_i = I_k \,, ~~~~~~~i \leq k 
\label{Identities} \,.
\eea

The above identities in particular state that $I_k^2 = I_k$, that is,
$I_k$ are idempotent, or in other words they are projection operators.
\par

By redefining the matrices $q_i$, $q_{i'}$ in the following way:
\begin{eqnarray}
  q_1 . I_3 & \rightarrow & q_1 \ \ , \ \    q_2 . I_1 \rightarrow q_2  \ \   , \ \   q_3 . I_2 \rightarrow q_3 \ ,
  \nonumber
  \\
  q_{1'} . I_2 & \rightarrow & q_{1'} \ \ , \ \    q_{2'} . I_3 \rightarrow q_{2'}  \ \   , \ \   q_{3'} . I_1 \rightarrow q_{3'} \ ,
\label{qs}
\end{eqnarray}
Eq. (\ref{TDI}) assumes the following form:
\begin{eqnarray}
  & (& - q_1 - q_{1'} + q_3 . D_2 + q_{2'} . D_3) . C_1
  \nonumber
  \\
  + & (& - q_2 - q_{2'} + q_1 . D_3 + q_{3'} . D_1) . C_2
        \nonumber
  \\        
  + & (& - q_3 - q_{3'} + q_2 . D_1 + q_{1'} . D_2) . C_3 = 0
        \label{TDIequations}
\end{eqnarray}
Since the three random processes $C_i \ , \ i=1, 2, 3$ are
independent, the above equation can be satisfied iff the three
matrices multiplying the three random processes are identically equal
to zero, i.e.:
\begin{eqnarray}
&  & - q_1 - q_{1'} + q_3 . D_2 + q_{2'} . D_3  = 0 \ ,
  \nonumber
  \\
&  & - q_2 - q_{2'} + q_1 . D_3 + q_{3'} . D_1 = 0 \ ,
 \nonumber
  \\        
&  & - q_3 - q_{3'} + q_2 . D_1 + q_{1'} . D_2  = 0 \ .
\label{TDIequations2}
\end{eqnarray}
Since the system of Eqs. (\ref{TDIequations2}) is identical in form to
the corresponding equations derived in \cite{TD2020} (see section 4.3
of \cite{TD2020} and equations therein), the solutions will assume the
same forms. It should be noticed, however, that the ``matrix''
expressions of the generators $\alpha, \beta, \gamma, \zeta$ can be
obtained from the usual TDI-expressions by taking into account that
the $qs$ had been redefined (see Eq.(\ref{qs}) ). This means that the
Sagnac combination $\alpha$, for instance, assumes the following form:
\begin{equation}
\alpha = (y_1 +  D_3 y_2 + D_1 D_3 y_3) -
(y_{1'} + D_2 y_{3'} + D_1 D_2 y_{2'}) \ ,
\label{alpha}
\end{equation}
where we have accounted for the identities given by
Eqs. (\ref{Identities}).  When considering seven time-samples of the
six one-way measurements, the above expression for $\alpha$ reduces to
the following vector:
\begin{equation}
  \alpha = 
\left(
    \begin{array}{c}
        0 \\
        0 \\
        -y_{1'}(t_2) - y_{3'}(t_0)\\
        y_1(t_3) - y_{1'}(t_3) + y_2(t_0) - y_{2'}(t_0) - y_{3'}(t_1)  \\
        y_1(t_4) - y_{1'}(t_4) + y_2(t_1) - y_{2'}(t_1) + y_3 (t_0) - y_{3'}(t_2) \\
        y_1(t_5) - y_{1'}(t_5) + y_2(t_2) - y_{2'}(t_2) + y_3 (t_1) - y_{3'}(t_3) \\
        y_1(t_6) - y_{1'}(t_6) + y_2(t_3) - y_{2'}(t_3) + y_3 (t_2) - y_{3'}(t_4) \\
        \vdots
    \end{array}
\right)
\label{alphaM}
\end{equation}
As in the case of the combination $X$ presented in the previous
section, here also the first few entries of the vector cannot cancel
the laser noises. This is because some of the measurements at those
time stamps are equal to zero. However, it is easy to verify that all
measurements at row seven and higher are different from zero and
reproduce the usual TDI combination $\alpha$ that cancels the laser
noise.

\section{TDI and matrix representations of delay operators}
\label{represent}

In this section we start with the general discussion of the algebraic
structure of time-delay operators and then go on to discuss the
homomorphism between the rings of time-delay operators and
matrices. We consider various cases of (i) time-delays which are
integer multiples of the sampling interval, (ii) the continuum case,
(iii) fractional time-delays with Lagrange interpolation and further
argue how the homomorphism could be extended to the situation of
time-dependent arm-lengths in which case the ring of delay operators
becomes non-commutative.
\par

We remark that the homomorphism concept is fundamental and should hold in every
situation of time delays; whether they are integer multiples of the
sampling interval, or fractional or time dependent. We argue in this section that
this is indeed so.

\subsection{General discussion of group and ring structures of time-delay operators}
\label{general}

Let us consider the data $y_j (t)$ as above. For the purpose of this
section we will drop the subscript from $y_j$ and call it just
$y (t)$. Also in the beginning of this section for purposes of
argument, we consider $- \infty < t < \infty$, that is $t \in \R$, the
set of real numbers. Later we will consider the realistic situation of
finite length data segment. A time delay operator $\D$ with delay $l$
acts on the $y$ as follows:

\bea
\D: \R &\longrightarrow & \R \,\no \\
y (t) & \longrightarrow & y (t - l) \,.
\eea

After having defined the delay operator $\D$, we may analogously
define several delay operators $\D_1, \D_2, ...$ with time delays
$l_1, l_2, ...$ respectively. The $\D$ operators are translations in
one dimension. The group operation here is then defined as the successive
application of the operators:
$$\D_1 \D_2 y (t) = \D_1 y (t - l_2) = y (t - l_1 - l_2). $$ 
With the operation so defined the $\D$s form an uncountable infinite
group. When the $l_1, l_2$ are constants, the group is Abelian and
coincides with the usual translation group in one dimension.
\par

Now consider the case of time-dependent arm-lengths. Then $l_1$ and
$l_2$ are functions of time themselves, and the product operation
becomes: \be \D_1 \D_2 y (t) = \D_1 y (t - l_2 (t)) = y [t - l_1 - l_2
(t - l_1)] \,, \ee which is in general non-commutative and the group
is non-Abelian. Then this is not the usual translation group, but
nevertheless it is a group, when the time rate of change of
arm-lengths respects relativity, that is, ${\dot l} < 1$.  Then any
$\D$ defines a bijective map from $\R$ to $\R$, so that the inverse
exists.
\par

When we consider several data streams $y_j$ as in two arm or three arm
interferometers, the $\D$ operators in fact form a polynomial ring
instead of only a group with the different $\D_j$ operators as
indeterminates \cite{DNV02}. The ring could be commutative or
non-commutative according as the arm-lengths are time independent
\cite{AET99,DNV02,NV04} or time dependent
\cite{STEA03,TEA04,SVDsymp,DNV10,TD2020}. The TDI data combinations
constitute a module over the polynomial ring of delay operators known
as the first module of syzygies. See \cite{DNV02,TD2020} for
details. The ring operations in general are defined in the obvious way
on a data stream $y (t)$. Given two operators $\D_1$ and $\D_2$: \bea
(\D_1  + \D_2) y (t) &=& \D_1 y (t) + \D_2 y (t) = y (t - l_1) + y (t - l_2) \, \no \\
\D_1 \D_2 y &=& = \D_1 y(t - l_2 (t)) = y [t - l_1 - l_2 (t - l_1)]
\,.  \eea

These operations can be extended to the whole ring by linearity. In
the examples in this paper, we consider the arm lengths to be constant
in time and so $\D_1 \D_2 = \D_2 \D_1$ and the polynomial ring is
commutative.

\subsection{Matrix representations of time-delay operators: integer valued time-delays}

We treat this case first as it is the easiest to understand
intuitively. We consider the more realistic situation where the data
segment is of finite duration $[0, T]$. We will also assume that the
data are sampled uniformly with sampling time interval $\Delta t$. Now
there are finite number of samples $N$ labeled by the times
$t_k = k~\Delta t, ~ k = 0, 1, 2, ..., N - 1$ and also we have
$N \Delta t = T$. See \cite{NumericalRecipes} for more details. Here
typically $N$ could be a large number, but the point is that it is
{\it finite}. So the measurements $y$ or the laser noise $C$ can be
represented by $N$ dimensional vectors in $\R^N$. Because noise is a
random process these are random vectors.
\par

The operators $\D$ now take the form of linear transformations from
$\R^N \longrightarrow \R^N$ and hence in our formulation can be
represented by $N \times N$ matrices which now for this case we will
represent by just $D$. With the arm-lengths taken as in section
\ref{SecIII}, the operators $\D_1, \D_2, \D_3$ are represented by the
matrices given by Eqs. (\ref{op1}), (\ref{op2}), (\ref{op3}). We have
essentially discretized the previous situation of the continuum.  In
the matrix representation we have represented the abstract TDI
operators $\D$ by the matrices $D$. The operations which were valid in
the abstract case map faithfully to the discretized version. The sum
and product of the $\D$ operators maps to the sum and product of the
$D$ matrices - the ring operations are preserved. This is in fact
known as a representation of a group or a ring in the literature. We
now formally define a representation \cite{Gel}:
\begin{mydef}
  Let $\G$ be a group and $V$ be a finite dimensional vector
  space. For every $g \in G$ there is associated
  $T_g: V \longrightarrow V$ a linear map. Then the map: \bea
  \varphi: \G & \longrightarrow & Hom (V, V) \, \no \\
  g & \longrightarrow & T_g \, \no \\
  \eea is called a representation if $\varphi$ is a group
  homomorphism, i. e. for every
  $g_1, g_2 \in \G, ~T_{g_1 g_2} = T_{g_1} T_{g_2}$ and the group
  identity $e \in G$ maps to $T_e = I$, the unit matrix. $Hom (V, V)$
  is the space of linear transformations from $V \longrightarrow V$ -
  the endomorphisms of $V$.
\end{mydef}

This definition easily extends to that of rings with identity, where
now the homomorphism must be a ring homomorphism; both operations of
the ring must be preserved under the homomorphism \cite{Burrow}. $V$
is called the carrier space. In our situation, $\varphi$ maps
$\D \longrightarrow D$ or $\varphi (\D) = D$; the delay operator $\D$
is mapped to the matrix $D$. It is easy to verify that this is indeed
a ring homomorphism. $V = \R^N$, the space of the vectors $y$ or $C$,
plays the role of the carrier space.
\par

We now elucidate the above discussion with an example of a TDI
observable for LISA. Considering a simple model of LISA with just
three time-delay operators $\D_1, \D_2, \D_3$ and constant
arm-lengths, any TDI observable is six component polynomial vector in
the delay operators. Let us consider the simplest of the TDI
observables, namely, $\zeta$. In the operator picture, it is an
element of the module of syzygies:

\be
\zeta = (-
\D_1, - \D_2, - \D_3, \D_1, \D_2, \D_3) \,.
\ee

In the matrix formulation, under the ring homomorphism, the matrix
form of $\zeta$ is:

\be \zeta = - D_1 y_1 - D_2 y_2 - D_3 y_3 + D_1 y_{1'} + D_2 y_{2'} +
D_3 y_{3'} \,, \ee where now the $D_k$ are $N \times N$ matrices and
the $y_j, ~y_{j'}$ are $N$ dimensional column vectors. $\zeta$ is now
a $N \times 1$ column vector which is devoid of laser frequency
noise. Let us now check whether $\zeta$ as defined here cancels the
laser frequency noise. We may write $\zeta$ in terms of the laser
noises $C_1, C_2, C_3$ from Eq. (\ref{oneways}): \be \zeta = D_1 (I_3
- I_2) C_1 + D_2 (I_1 - I_3) C_2 + D_3 (I_2 - I_1) C_3 \,.  \ee At the
time $t_k$, we then have, \be \zeta (k) = (I_3 - I_2) C_1 (k - 1) +
(I_1 - I_3) C_2 (k - 2) + (I_2 - I_1) C_3 (k - 3) \,, \ee where in
order to avoid clutter we have written $k$ for the sampling time
$t_k$. From the above equation we deduce that $\zeta (k) = 0$ for
$k \geq 5$ and also $\zeta (0) = \zeta (1) = \zeta (2) = 0$. However,
$\zeta (3) = C_2 (1) - C_1 (2) \neq 0$ and
$\zeta (4) = C_3 (1) - C_2 (2) \neq 0$ and so at these sampling times
the laser noise does not cancel.
\par
 
In the algebraic approach \cite{DNV02, TD2020}, any TDI observable is
a 6-tuple polynomial vector in the operators $\D_k$. In the matrix
formulation, since the operators $\D_k$ map to the matrices $D_k$
under the ring homomorphism, an operator polynomial maps also to a
matrix. Thus in the matrix formulation any TDI observable is expressed
in terms of 6 matrices $q_i, q_{i'}$; the polynomials $q_i,~q_{i'}$ in
the operators $\D_k$ are now interpreted as matrices. In the two arm
configuration discussed in Section \ref{SecII} only two matrices are
required $A_1$ and $A_2$. In terms of the $D_k$ matrices defined,
$A_1 = D_1 - I_1$ and $A_2 = D_2 - I_2$. In a recent work
\cite{Vallisneri2020} the two $N \times N$ matrices are juxtaposed in
the form of a $2 N \times N$ matrix $M$, called the design matrix, and
in which the $y_1, y_2$ measurements are interleaved together in rows
as in Eq. (\ref{design}).  Note that the TDI combinations presented in
matrix form can be repackaged in a format $T . y$
\cite{Vallisneri2020}, which might turn out to be more advantageous
for numerical manipulations and data analysis.
\par

A Bayesian inference approach has been adopted by Romano and Woan
\cite{PCA2006}.  They set up a noise covariance matrix of the data
streams $y_j, y_{j'}$ and perform a principal component analysis. From
the principal components they identify large eigenvalues with laser
noise and so distinguish it from the signal. We remark that this is
also a matrix representation of the original TDI, although a little
more complex - it is a tensor representation or product representation
\cite{Gel}. The covariance matrix is a second rank tensor. Any entry
of the covariance matrix $C_{ik}$ is an ensemble average of outer
products of the form
$y_\a (i) y_{\b}^T (k), ~{y}_{\a'} (i) y_{\b}^T (k)$ or
${y}_{\a'} (i) {y}_{\b'}^T (k)$. We use Greek indices
$\a, \b = 1, 2, 3$ to label data streams and operators to distinguish
them from time samples which are tensor indices.  At each $(i, k)$,
the products in $y$ contain tensor products - for example,
$y_1 (i) {y}_{1'}^T (k)$ contains products of the $D$ matrices,
namely, $D_{3 ij} D_{2 mk} C_{2j} C_{3m}$. The outer product of the
vectors $C_{2}$ and $C_{3}$, namely,
$C_{2} \otimes C_{3} \in \R^N \otimes \R^N$ is a tensor of second rank
and the product of the $D$s acts on this tensor. These products of
$D$s define the tensor representation.  $\R^N \otimes \R^N$ acts as
the carrier space for this representation.

\subsection{The continuum case}

From a logical point of view, this case could have been addressed
immediately after subsection \ref{general}, but for concreteness sake,
we felt that we should first deal with the easier case of constant
time-delays which are integer multiples of the sampling interval.

We have already shown that homomorphism holds for the case of integer
multiples of sampling interval. As a matter of principle, one may
argue that if the Doppler data could be sampled at a rate as high as
required by TDI (corresponding to a sampling time of about (10 m/c)
sec) then we may approach the previous case of integer valued
time-delays and the equality $\phi(\D_1 \D_2) = \phi(\D_1) \phi(\D_2)$
would seem to hold. So this motivates us, on a theoretical basis (also
it is instructive), to examine this question by taking the continuum
limit of the sampling interval $\Delta t \longrightarrow 0$. Then the
matrix representation of a delay operator $\D_1$ with delay $l_1 (t)$
tends to a delta function
$\delta [t' - (t - l_1 (t))] \equiv D_1 (t, t')$. Here the matrix
$D_1 (t, t')$ - a function of 2 variables - acts on the continuous
data stream $y (t)$ as follows: \be \D_1 y (t) = \int dt'~ D_1 (t, t')
y (t') = \int dt'~ \delta [t' - (t - l_1 (t))]~ y (t') ~=~ y (t - l_1
(t)) \,, \ee which is consistent with the usual definition of the
operator $\D_1$. Here the homomorphism $\phi$ is
$\phi (\D_1) = D_1 (t, t')$. If one takes two such operators even with
time-dependent delays $l_1 (t)$ and $l_2 (t)$, and applies the two
operators successively then the result is again a delta function with
a delay $l_1 (t) + l_2 (t - l_1 (t))$ as shown below: \bea
\phi (\D_1) \star \phi (\D_2) &=& (D_1 \star D_2)~(t, \tpp) \,, \no \\
&=& \int dt'~ D_1 (t,t') ~ D_2 (t', \tpp) \,, \no \\
&=& \int dt'~\delta [t' - (t - l_1 (t))]~\delta [\tpp - (t' - l_2 (t'))] \,, \\
&=& \delta [\tpp - \{t - l_1 (t) - l_2 (t - l_1 (t)) \}] \equiv \phi
(\D_1 \D_2) \,.  \eea This proves that the matrix representation in
the continuum case is also a homomorphism. In general, \be \phi (\D_1)
\star \phi (\D_2) = D_1 \star D_2 \neq D_2 \star D_1 = \phi (\D_2)
\star \phi (\D_1).  \ee The operators do not commute in general when
the arm lengths are time dependent. The operators then form a
non-commutative polynomial ring.  When the delays are constants, the
operators $\D_1$ and $\D_2$ commute and the operators form a
commutative polynomial ring. So far we have shown that the
homomorphism holds in the continuum limit in addition to the case of
delays being integer multiples of the sampling interval (constant
time-delays) - the opposite end, so to speak.

\subsection{Fractional time-delays and time-dependent arm-lengths}

In practice one has nonzero sampling intervals $\Delta t > 0$. But for
LISA, because of practical limitations, this sampling would be too
coarse to be used in the TDI algorithms to cancel the laser frequency
noise. For this purpose one would require data at points between the
sample points. One then applies appropriate fractional delay filters
to the Doppler measurements to achieve this goal digitally.
Fractional delays may be implemented using an interpolation
scheme. Here we employ Lagrange interpolation as in
\cite{Vallisneri2020}.  We consider three cases: 

\begin{enumerate} 
\item Single interval for all delays, 
\item Different intervals for each delay,
\item Time-dependent delays. 
\end{enumerate}

\subsubsection{Single interval for all delays (time-independent)}

Without loss of generality, we consider $m$ sample points $t = 0, 1, ..., m - 1$ with $\Delta t = 1$. We denote this interval by $I_0 = \{0, 1, 2, ..., m - 1 \}$ which accommodates all delays. The interpolation operation can be cast in a matrix form with a matrix acting on the
data. More specifically, one can envisage a $m \times m$ matrix of
Lagrange polynomials $\D(\alpha)$, where $\alpha$ is the delay, acting
on the data $y$. We write the delays as $\a, \b, ...$ in order to not
confuse with the Lagrange polynomials which are also denoted by
$l_i$. We consider two delays $\a$ and $\b$ with the corresponding $m \times m$ matrices $\D (\a)$ and $\D (\beta)$. To establish the homomorphism, we show that $\D (\alpha + \beta) = \D (\alpha) \D (\beta)$. This result easily follows from the properties of Lagrange polynomials, namely, the addition theorem for
Lagrange polynomials. We give the proof of the addition theorem in the
appendix \ref{app:add_thm} .

For concreteness, consider just $m = 3$ points at $t = 0, 1, 2 $. Then the matrix $\D (\a)$ is:
\be
\D (\a) =  \left \| \begin{array}{ccc} 
l_0 (\a) & l_1 (\a) & l_2 (\a) \\
l_0 (\a + 1) & l_1 (\a + 1) & l_2 (\a + 1) \\
l_0 (\a + 2) & l_1 (\a + 2) & l_2 (\a + 2) 
\end{array} \right \| \equiv D_{jk} (\a) \,, 
\ee
where $D_{jk} (\a) = l_k (\a + j)$. Taking two such matrices corresponding to $\a$ and $\b$ and multiplying them together, we have,
\be
\sum_{k} D_{jk} (\a) D_{kn} (\b) = \sum_{k} l_k (\a + j) l_n (\b + k) \equiv l_n (\a + \b + j) = D_{jn} (\a + \b) \,,
\label{hom}
\ee where we have used the addition theorem in the appendix
\ref{app:add_thm}. Although we have just used 3 time stamps the
results are generally true for $m$ points. Also one might think, that
since the product of Lagrange polynomials appears as entries in the
product of the matrices, it might lead to polynomials of degree
$2m - 2$. But this does not happen, as the addition theorem shows; the terms of degree greater
than $m - 1$ cancel out, leaving behind a $m - 1$ degree polynomial.

\subsubsection{Different intervals for each delay (time-independent)}

In practice, choosing the same set of sample points may not be
feasible for delays much greater than the sampling interval and so
different sets of sample points must be chosen for different delays
but then the matrices may {\it appear} different, because the Lagrange
polynomials are translated. But then care must be taken to translate
the matrices to a common reference in order to compare them. Then the
closure property of the polynomials can be explicitly seen to hold. We
may see this as follows:

Let $I_r = \{r, r + 1, ..., r + m -1 \}$ be the interpolation
interval containing $m$ points around $\a$ and a corresponding
interval $I_s = \{s, s + 1, ..., s + m - 1 \}$ around $\b$. Let
$l_j (t), ~j = 0, 1, ..., m - 1$ be the Lagrange polynomials for the
reference interval $I_0 = \{0, 1, 2, ..., m - 1 \}$. We will call
these the {\it basic} Lagrange polynomials referred to $t = 0$. Then
the Lagrange polynomials for the interval $I_r$ are just the
translated versions of $l_j (t)$, namely, $l_j (t - r)$ and similarly
$l_j (t - s)$ for $I_s$. In this case the translated matrix
representation is $D^{(r)}_{jk} (\a) = l_k (\a - r + j)$ for delay
$\a$ and $D^{(s)}_{jk} (\b) = l_k (\b - s + j)$ for $\b$. Now the
total delay $\a + \b$ in general will lie between $r + s$ and
$r + s + 2m - 2$. We may choose $r$ and $s$ so that
$\a + \b \leq r + s + m - 1$ so that the relevant interval is $I_{r + s} = \{r + s, r + s + 1, ..., r + s + m - 1 \}$. The homomorphism is given by: \be
\sum_{k} D^{(r)}_{jk} (\a) D^{(s)}_{kn} (\b) = \sum_{k} l_k (\a - r +
j) l_n (\b - s + k) \equiv l_n (\a + \b - (r + s) + j) = D^{(r +
  s)}_{jn} (\a + \b) \,.
\label{hom2}
\ee 
This establishes the homomorphism for this case. We have again appealed to the addition theorem of Lagrange
polynomials proved in appendix \ref{app:add_thm}. We could have
perhaps argued that here, since we are concerned about matters of
principle, we may have chosen $m$ sufficiently large to cover all
delays. But we preferred to explicitly establish the homomorphism for
the case of each delay with different interpolating intervals.

\subsubsection{Time-dependent delays}

We further add that Eq. (\ref{hom}) is valid for
time dependent delays also. Now both $\a$ and $\b$ become functions of
time. If one applies the delay $\b$ first and then $\a$, the combined
delay is $\a + \b (\a) \equiv \a \oplus \b$ and in the reverse case it
is $\b + \a (\b) = \b \oplus \a$ which are in general unequal. Then we
have the situation: \be \D (\a \oplus \b) = \D [\a + \b (\a)] = \D
(\a) \D [\b (\a)] \neq \D (\b) \D [\a (\b)] = \D [\b + \a (\b)] = \D
(\b \oplus \a) \,.
\label{hom_timedep}
\ee Eq. (\ref{hom_timedep}) shows that the homomorphism also holds for
time-dependent fractional delays for which the operators do not
commute in general.
\par

In summary, we emphasize here that the matrix formulation is a ring
representation of the original TDI formulation. In principle it is no
different. However, in practice, there may be advantages to this
approach, because representations using matrices lend themselves to
easy analytical and numerical manipulations.

\section{Conclusions}
\label{SecIV} 

The main result of our article has been the demonstration that the
delay operators characterizing TDI may be represented as
matrices. Through this approach we recovered the well known result
characterizing TDI: the cancellation of the laser noise in an
unequal-arm interferometer is a ``local operation'' as it is achieved
at any time $t$ by linearly combining only a few neighboring
measurement-samples. Our conclusion is the consequence of correctly
accounting for the time-mismatch between the arrays of the Doppler
  measurements and that of the laser noise.

In mathematical terms, we have shown that the cancellation of the
laser noises using matrices is just the ring representation of the
original TDI formulation and it is not different from it. We
mathematically prove the homomorphism between the delay operators and
their matrix representation holds in general. We have covered 
all cases of interest: (i) time-delays that are constant integer multiples of the sampling interval, (ii) the continuum limit $\Delta t \longrightarrow 0$
including time-dependent arm-lengths and (iii) fractional time-delays when arm-lengths are time-independent (same interval and different intervals of interpolation) or time-dependent. For the fractional delay filters, Lagrange
interpolation has been used to establish the homomorphism.

It should be said, however, that the matrix approach we have
introduced might offer some advantages to the data processing and
analysis tasks of currently planned gravitational wave missions
\cite{LISA2017,Taiji,TianQin} as it is more flexible, allows for
easier numerical implementation and manipulation and also adapts to
time-dependent arm-lengths in a natural way.  Further on another
front, it might in fact be possible to extend to our matrix
formulation of TDI the data processing algorithm discussed in
\cite{Vallisneri2020} to handle data gaps. We have just started to
analyze this problem and might report about its solution in a
forthcoming article.

We remark that regardless of the approach we follow, both the original
as well as the matrix approaches look for null spaces whose vectors
describe the TDI observables. In the original TDI approach, the first
module of syzygies is in fact a null space - the kernel of a
homomorphism; the kernel is important because it contains elements,
namely, those TDI observables that map the laser noise to zero.

\section*{Acknowledgments}

M.T. thanks the Center for Astrophysics and Space Sciences (CASS) at
the University of California San Diego (UCSD, U.S.A.) and the National
Institute for Space Research (INPE, Brazil) for their kind hospitality
while this work was done. S.V.D. acknowledges the support of the
Senior Scientist Platinum Jubilee Fellowship from NASI, India.

\appendix
\section{Addition theorem for Lagrange polynomials}
\label{app:add_thm}

First, consider just $m = 3$ points at $t = 0, 1, 2$ and let
$l_j (t), ~~j = 0, 1, 2$ be the Lagrange polynomials. We do not need
them explicitly. Let $p (t)$ be the interpolating polynomial which is
required to pass through the points $y_0, y_1, y_2$ at $t = 0, 1, 2$
respectively. Then we have, \be p (t) = l_0 (t) y_0 + l_1 (t) y_1 +
l_2 (t) y_2 \,.  \ee We just need to use the property of Lagrange
polynomials: \be l_j (t = k) = \delta_{jk} \ee From this we have
$p (k) = y_k$ and so: \be p (t) = l_0 (t) p(0) + l_1 (t) p (1) + l_2
(t) p (2) \,.
\label{interpol}
\ee Consider the first term of the product matrix, namely,
$l_0 (\a + \b)$ and set it equal to $p (\a)$, where now $\b$ plays the
role of a constant. Then at each value of $\a = 0, 1 , 2$ we have
$p (k) = l_0 (\b + k)$. Thus from Eq. (\ref{interpol}) we obtain: \be
l_0 (\a + \b) = \sum_{k = 0}^2 l_k (\a) l_0 (\b + k) \,.  \ee In
general for $m$ time samples and the $n^{\rm th}$ Lagrange polynomial
$l_n$ we have: \be l_n (\a + \b) = \sum_{k = 0}^{m - 1} ~ l_k (\a) l_n
(\b + k) \,.
\label{eq:add}
\ee This is the addition theorem for Lagrange polynomials for integer
valued nodes at $k = 0, 1, ..., m$.

\bibliographystyle{apsrev}
\bibliography{refs}
\end{document}